\documentclass[twocolumn]{aastex631}
\usepackage{CJK}
\usepackage{amsmath}
\usepackage{xcolor}
\usepackage{bm}
\usepackage{soul}

\shorttitle{ion and electron temperatures in the solar wind}
\shortauthors{Shi et al.}
\graphicspath{{./}{figures/}}

\begin{document}
\begin{CJK*}{UTF8}{gbsn}
\title{Proton and electron temperatures in the solar wind and their correlations with the solar wind speed}

\correspondingauthor{Chen Shi}
\email{cshi1993@ucla.edu}

\author[0000-0002-2582-7085]{Chen Shi (时辰)}
\affiliation{Department of Earth, Planetary, and Space Sciences, University of California, Los Angeles \\
Los Angeles, CA 90095, USA}

\author[0000-0002-2381-3106]{Marco Velli}
\affiliation{Department of Earth, Planetary, and Space Sciences, University of California, Los Angeles \\
Los Angeles, CA 90095, USA}
    
\author[0000-0001-9231-045X]{Roberto Lionello}
\affiliation{Predictive Science Inc \\
San Diego, CA 92121, USA}

\author[0000-0002-1128-9685]{Nikos Sioulas}
\affiliation{Department of Earth, Planetary, and Space Sciences, University of California, Los Angeles \\
Los Angeles, CA 90095, USA}

\author[0000-0001-9570-5975]{Zesen Huang (黄泽森)}
\affiliation{Department of Earth, Planetary, and Space Sciences, University of California, Los Angeles \\
Los Angeles, CA 90095, USA}
    
\author[0000-0001-5258-6128]{Jasper S. Halekas}
\affiliation{Department of Physics and Astronomy, University of Iowa\\
Iowa City, IA 52242, USA}

\author[0000-0003-2880-6084]{Anna Tenerani}
\affiliation{Department of Physics, The University of Texas at Austin, \\
     TX 78712, USA}
     
\author[0000-0002-2916-3837]{Victor R\'eville}
\affiliation{IRAP, Universit\'e Toulouse III - Paul Sabatier,
CNRS, CNES, Toulouse, France}

\author{Jean-Baptiste Dakeyo}
\affiliation{LESIA, Observatoire de Paris, Universit e PSL, CNRS, Sorbonne Universit e, Universit e de Paris, 5 place Jules Janssen, 92195 Meudon, France}

\author[0000-0001-6172-5062]{Milan Maksimovi\'c}
\affiliation{LESIA, Observatoire de Paris, Universit e PSL, CNRS, Sorbonne Universit e, Universit e de Paris, 5 place Jules Janssen, 92195 Meudon, France}

\author[0000-0002-1989-3596]{Stuart D. Bale}
\affil{Physics Department, University of California, Berkeley, CA 94720-7300, USA}
\affil{Space Sciences Laboratory, University of California, Berkeley, CA 94720-7450, USA}



\begin{abstract}
The heating and acceleration of the solar wind remains one of the fundamental unsolved problems in heliophysics. It is usually observed that the proton temperature $T_i$ is highly correlated with the solar wind speed $V_{SW}$, while the electron temperature $T_e$ shows anti-correlation or no clear correlation with the solar wind speed. 
Here we inspect both Parker Solar Probe (PSP) and WIND data and compare the observations with simulation results. PSP observations below 30 solar radii clearly show a positive correlation between proton temperature and wind speed and a negative correlation between electron temperature and wind speed. One year (2019) of WIND data confirm that proton temperature is positively correlated with solar wind speed, but the electron temperature increases with the solar wind speed for slow wind while it decreases with the solar wind speed for fast wind. Using a one-dimensional Alfv\'en-wave-driven solar wind model with different proton and electron temperatures, we for the first time find that if most of the dissipated Alfv\'en wave energy heats the ions instead of electrons, a positive $T_i-V_{SW}$ correlation and a negative $T_e-V_{SW}$ correlation arise naturally. If the electrons gain a small but finite portion of the dissipated wave energy, the $T_e-V_{SW}$ correlation evolves with radial distance to the Sun such that the negative correlation gradually turns positive. The model results show that Alfv\'en waves are one of the possible explanations of the observed evolution of proton and electron temperatures in the solar wind.
\end{abstract}



\section{Introduction} \label{sec:intro}
Solar wind is the plasma flow ejected from solar corona, filling the interplanetary space. It carries a large amount of mass and energy out of the Sun and serves as the medium of various physical processes and structures, such as waves and turbulence, coronal mass ejection, and magnetic reconnection. Solar wind continuously interacts with the Earth and injects energy in the Earth's magnetosphere, causing strong disturbances of the Earth's magnetosphere. Understanding the generation of solar wind and the dynamics of the plasma in the solar wind is necessary not only for a better space weather prediction but also for a deeper insight of the fundamental plasma astrophysics.


\begin{figure*}
    \centering
    \includegraphics[width=\hsize]{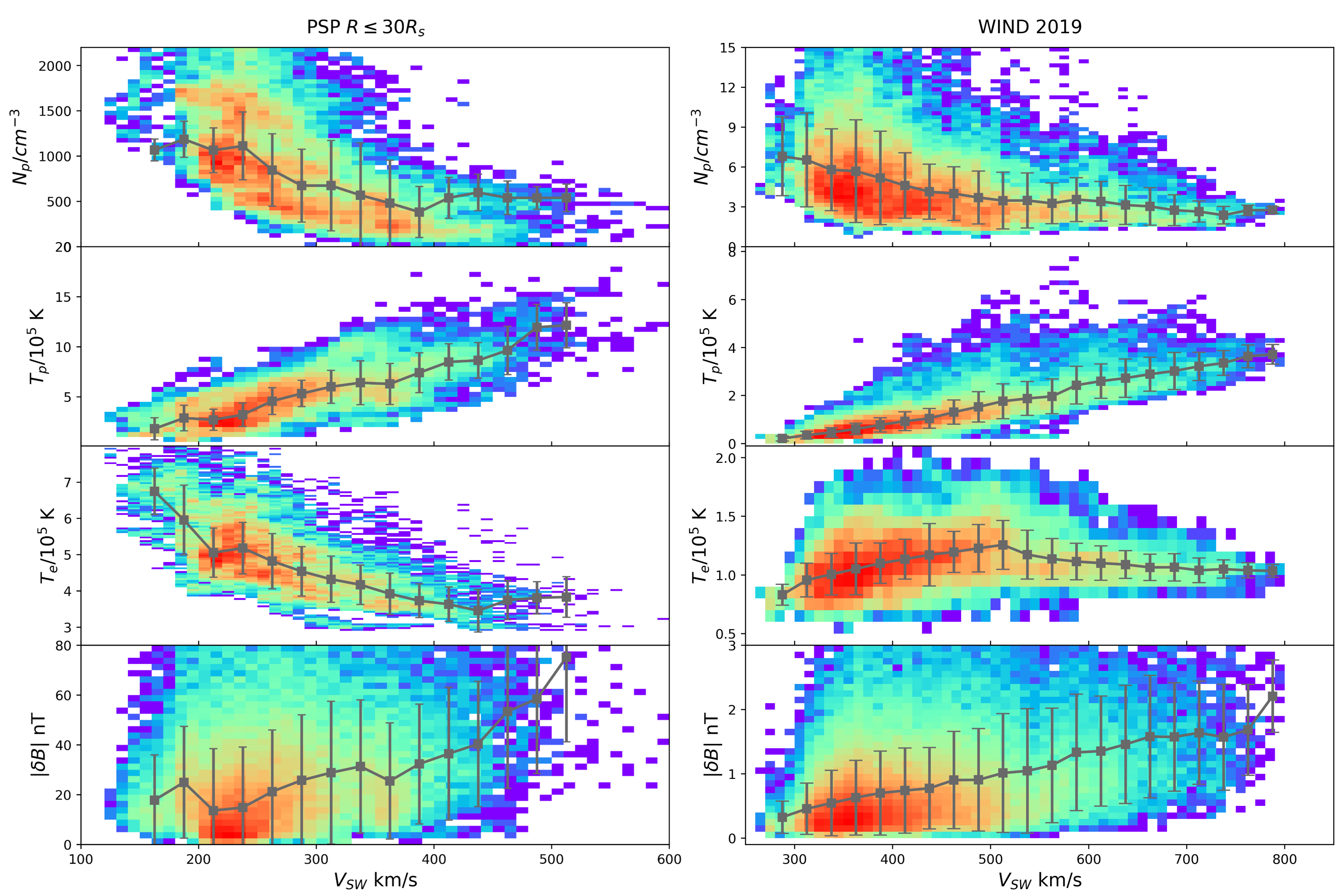}
    \caption{Left column: Distribution of data collected below 30 solar radii during first nine orbits of Parker Solar Probe. Right column: Distribution of one year (2019) of data from WIND. From top to bottom, the rows are proton number density $N_p$, proton temperature $T_p$, electron temperature $T_e$, and amplitude of the magnetic field fluctuations $\left| \delta \bm{B} \right|$ as functions of the radial solar wind speed $V_{SW}$. For PSP data, we use one-minute time windows to calculate the average values of the quantities and the fluctuation strength. For WIND data, we use five-minute time windows. The gray squares show the median values and the error bars show the root-mean-squares of the binned data.}
    \label{fig:insitu_observation}
\end{figure*}

A significant amount of energy is needed for the plasma to escape the solar gravity and to accelerate to a supersonic speed. Early works \citep{parker1958dynamics,parker1964dynamicalI,parker1964dynamicalII,parker1965dynamical} show that an isothermal solar corona with strong thermal conduction can generate the observed solar wind speeds. However, in-situ measurements imply that in the inner heliosphere both ion and electron temperatures of the solar wind decay with radial distance to the Sun and the radial profiles of the temperatures can be fitted with polytropic relations \citep{marsch1982solar,richardson1995radial,vstverak2015electron,boldyrev2020electron,shi2022acceleration}. The polytropic indices deduced from the in-situ measurements are in general smaller than the adiabatic index $5/3$, implying that in-situ heating mechanisms are important. One possible source of the in-situ heating is the turbulence energy cascade. In the solar wind, predominantly outward propagating Alfv\'en waves exist \citep{belcher1971large}, while smaller-amplitude inward propagating Alfv\'en waves are generated due to processes including wave reflection induced by the gradient of Alfv\'en speed \citep{heinemann1980non} and the large-scale stream shears \citep{roberts1992velocity,shi2020propagation}. The nonlinear interaction between the outward and inward propagating waves leads to a turbulence energy cascade \citep{kraichnan1965inertial}, which eventually dissipates the energy of the electromagnetic fields into the plasma energy through wave-particle interactions \citep{kasper2013sensitive,kobayashi2017three} and intermittent structures \citep{osman2012intermittency,matthaeus2015intermittency,sioulas2022statistical}. 

It has been long observed that in the solar wind, the proton temperature has a strong positive correlation with the solar wind speed \citep{burlaga1973solar,lopez1986solar,matthaeus2006correlation,demoulin2009temperature,elliott2012temporal,shi2021alfvenic,hofmeister2022area}. 
This positive correlation indicates the possibility that a particular mechanism contributes to the proton heating and solar wind momentum simultaneously. 
On the contrary, electron core temperature usually shows anti-correlation with the solar wind speed \citep{marsch1989cooling,halekas2020electrons,maksimovic2020anticorrelation} and this anti-correlation may originate at the source of solar wind due to interchange reconnection \citep{fisk2003acceleration,gloeckler2003implications}. 
In-situ measurements of the fractions between heavy ions, which are good proxies for freeze-in temperatures of the solar corona, reveal that the coronal electron temperature has an anti-correlation with solar wind speed \citep{geiss1995southern,ko1997empirical,gloeckler2003implications,von2011polar}.

Here, we propose a mechanism that the different temperature-speed correlations for protons and electrons are possibly generated in-situ as the solar wind propagates. The major factor is how the energy source for the acceleration of solar wind deposits differently into protons and electrons. 
As the turbulence energy cascades from MHD inertial scales toward ion kinetic scales, various kinetic processes may arise and they determine how the electromagnetic energy eventually heats the ions and electrons. Two major processes are kinetic Alfv\'en waves (KAWs) which mainly heat the electrons through Landau damping, and ion cyclotron waves (ICWs) which heat the ions through cyclotron resonance. Both KAWs \citep[e.g.][]{podesta2013evidence,salem2012identification} and ICWs \citep[e.g.][]{jian2010observations,jian2009ion} are identified in the solar wind, and recent observation made by Parker Solar Probe shows that the two modes may coexist \citep{huang2020kinetic}. Gyrokinetic theory and simulations show that the ratio between ion heating and electron heating during dissipation of Alfv\'enic turbulence is positively correlated with ion beta (ratio between ion thermal pressure and magnetic pressure) \citep{howes2008model,schekochihin2009astrophysical,howes2010prescription,kawazura2019thermal}. Moreover, existence of compressive component in the turbulence will increase the ratio between ion and electron heating \citep{kawazura2020ion}. Hybrid-kinetic simulations using realistic solar wind parameters at 1 AU show that 75-80\% of the cascaded turbulence energy heats the ions \citep{arzamasskiy2019hybrid}. Recent theoretical work has suggested that conservation of magnetic helicity prevents the turbulence energy from cascading toward sub-ion scales, thus most of the cascaded energy is absorbed by the ions \citep{squire2022high}. \citet{bacchini2022kinetic} show that during the transition from Alfv\'en waves to kinetic Alfv\'en waves at sub-ion scales, ions can gain more energy than electrons because the kinetic energy of the waves is mostly accessible to ions instead of electrons. In addition to the wave-particle interaction, other effects such as intermittency \citep[e.g.][]{osman2012intermittency} and stochastic heating \citep[e.g.][]{chandran2010perpendicular} may also contribute to the differential heating process. \citet{sioulas2022preferential}, using Parker Solar Probe measurements, show that protons can gain more energy from the intermittent structures than the electrons. The stochastic heating is positively correlated with the turbulence strength \citep{vech2017nature} and is shown to be significant throughout the inner heliosphere \citep{martinovic2019radial,martinovic2020enhancement}, contributing to the perpendicular temperature of ions.

In this study, we show that, for an Alfv\'en wave driven solar wind, if most of the wave energy dissipates into ions, the positive $T_p-V_{SW}$ ($T_p$ is proton temperature and $V_{SW}$ is solar wind speed) correlation and negative $T_e-V_{SW}$ ($T_e$ is electron temperature) correlation are naturally generated. The paper is organized as follows: In Section \ref{sec:observation}, we present Parker Solar Probe (PSP) and WIND observations of the solar wind and show the correlations between temperatures and solar wind speed. In Section \ref{sec:simulations}, we describe the 1D Alfv\'en wave driven solar wind model used in this study and present the simulation results. In Section \ref{sec:discuss} we discuss the underlying mechanism that explains the numerical results. In Section \ref{sec:conclusion} we conclude this study.

\section{WIND \& PSP observations}\label{sec:observation}
We use PSP and WIND data to investigate the correlation between the solar wind speed and various solar wind parameters. For PSP, we use data from the first nine orbits and we only select data collected below 30 solar radii. For proton measurements, we mainly use data from the electrostatic analyzer (SPAN-Ion) but use the Faraday cup (SPC) for the first orbit when high-quality SPAN-Ion data is unavailable \citep{fox2016solar,kasper2016solar}. The electron temperature is the derived core temperature by fitting the electron velocity distribution functions measured by SPAN-Electron \citep{halekas2020electrons}. The time cadence of SPAN data is typically 7-14 sec, and the time cadence of SPC data is around 0.44 sec. The magnetic field data is collected by the fluxgate magnetometer with a time cadence of 3.4 milliseconds \citep{fox2016solar,bale2016fields}. 
For WIND, we use one year of data collected in 2019 and we have verified that the 2020 data give very similar results. The proton data is from the 3D Plasma Analyzer (3DP) electrostatic analyzers with three second cadence \citep{lin1995three}, the electron data is from the Solar Wind Experiment (SWE) electron instruments with 6-12 second cadence \citep{ogilvie1995swe}, and the magnetic field data is from the Wind Magnetic Field Investigation (MFI) fluxgate magnetometers with three second cadence \citep{lepping1995wind}.

In Figure \ref{fig:insitu_observation}, we show distribution of PSP data on the left column and distribution of WIND data on the right column. From top to bottom rows are proton number density $N_p$, proton temperature $T_p$, electron temperature $T_e$, and amplitude of the magnetic field fluctuations $\left| \delta \bm{B} \right|$ as functions of the radial solar wind speed $V_{SW}$. For PSP data, we use one-minute time windows to calculate the average values of the quantities and the fluctuation strength, defined as the root-mean-square (RMS) of the magnetic field. For WIND data, we use five-minute time windows. The gray squares show the median values and the error bars show the RMS of the binned data. It is clear in both datasets that higher solar wind speed in general corresponds to lower density, higher proton temperature, and stronger magnetic field fluctuations. As previously shown in \cite{maksimovic2020anticorrelation,halekas2020electrons,salem2021precision,dakeyo2022statistical}, the electron temperature has a strong negative correlation with the solar wind speed as observed by PSP. However, at 1 AU, $T_e$ decreases with $V_{SW}$ for winds faster than about 520 km/s but seems to increase with $V_{SW}$ for slower wind streams.

\begin{figure*}
    \centering
    \includegraphics[width=\hsize]{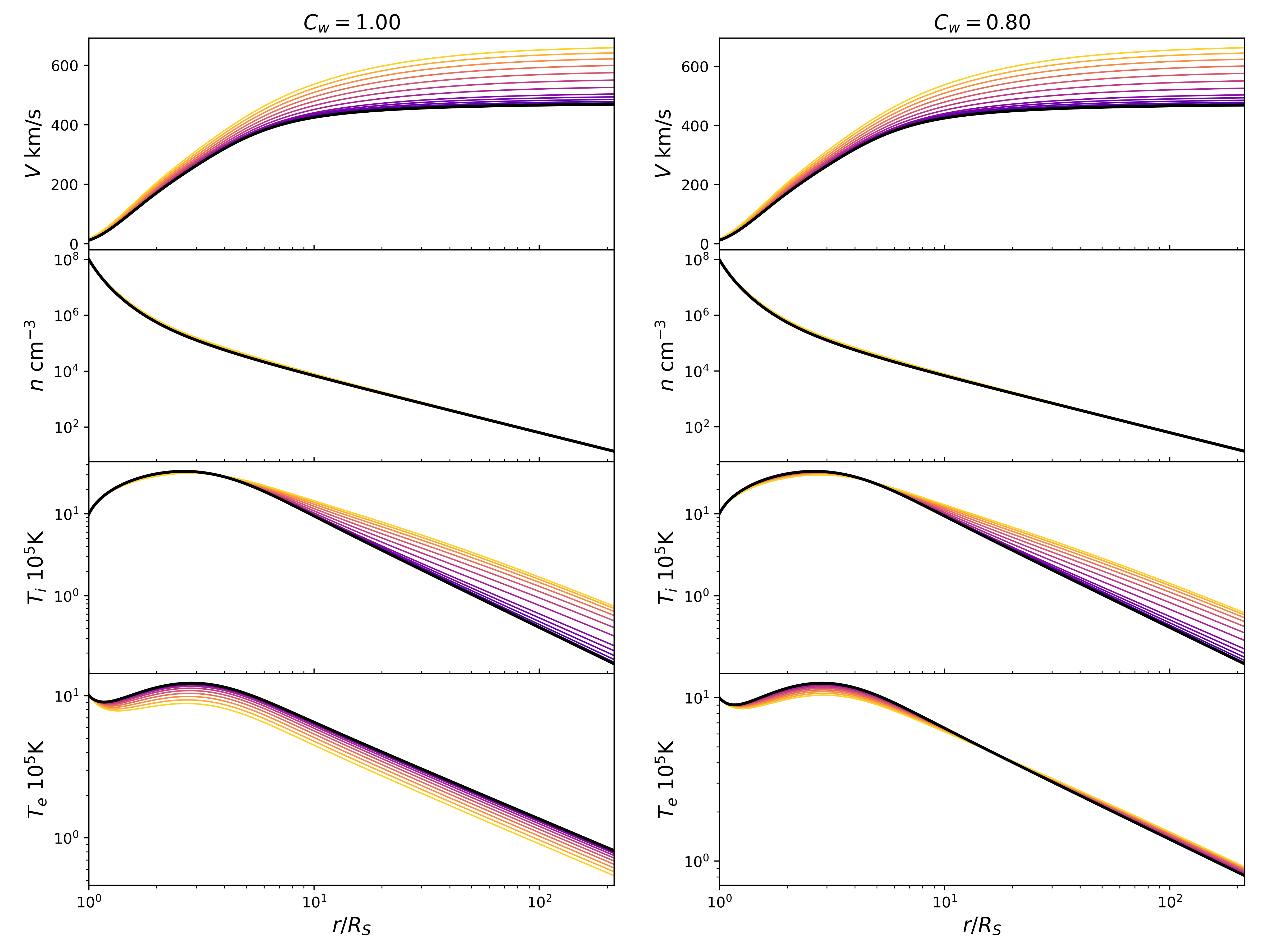}
    \caption{Radial profiles of various quantities in two sets of simulations. Left column has $C_w=1$, i.e. all the dissipated wave energy heats the ions. Right column has $C_w=0.8$, i.e. 20\% of the dissipated wave energy heats the electrons. From top to bottom rows are solar wind speed, 
    plasma number density, ion temperature, and electron temperature respectively. In each panel, curves with different colors correspond to different wave amplitudes at the inner boundary. From dark to light, colors correspond to increasing wave amplitudes of $[0,5,10,15,20,25,30,40,50,60,70,80,90,100]$ km/s.
    }
    \label{fig:radial_profile}
\end{figure*}

\section{1D two-temperature Alfv\'en wave powered solar wind model}\label{sec:simulations}
\subsection{Model description}
We utilize a 1D Alfv\'en-wave-driven solar wind model with different ion (proton) and electron temperatures. Wave-driven solar wind models have been developed and widely used to analyze the heating and acceleration of solar wind \citep[e.g.][]{cranmer2005generation,cranmer2007self,chandran2009alfven,verdini2007alfven,verdini2009turbulence,lionello2014validating,shoda2018self,reville2020role}. While most of the previous works assume a one-fluid solar wind, some works have adopted a two-fluid solar wind model with different proton and electron temperatures \citep{chandran2011incorporating,adhikari2022mhd}. The model used in the current study is very similar to the one-fluid model used by \citet{reville2020role} but with independent ion and electron temperatures. The model equations are
\begin{subequations}
\begin{equation}
        B(r) = \frac{A_0}{A(r)} B_0
    \end{equation}
    \begin{equation}
        \frac{\partial \rho}{\partial t} = - \frac{1}{A} \frac{\partial}{\partial r} \left( \rho V A\right)
    \end{equation}
    \begin{equation}
    \begin{aligned}
    \label{momentum}
        \frac{\partial V}{\partial t} =&  - V \frac{\partial V}{\partial r} - \frac{1}{\rho} \frac{\partial}{\partial r} \left( P + \frac{1}{2} \varepsilon \right)  -  \frac{GM}{r^2} 
    \end{aligned}
    \end{equation}
    \begin{equation}
    \begin{aligned}
        \frac{\partial P_i}{\partial t} = & - V \frac{\partial P_i}{\partial r} - \gamma_i \frac{1}{A} \frac{\partial \left(A V \right)}{\partial r}   P_i + (\gamma_i - 1) Q_i 
    \end{aligned}
    \end{equation}
    \begin{equation}\label{eq:pe}
    \begin{aligned}
        \frac{\partial P_e}{\partial t} = & - V \frac{\partial P_e}{\partial r} - \gamma_e \frac{1}{A} \frac{\partial  \left(A V \right)}{\partial r}  P_e + (\gamma_e - 1) Q_e 
    \end{aligned}
    \end{equation}
    \begin{equation}\label{eq:epsilon+}
    \begin{aligned}
        \frac{\partial \varepsilon^+}{\partial t} = & - (V+V_A) \frac{\partial \varepsilon^+}{\partial r} - \frac{1}{A} \frac{\partial }{\partial r} \left(A \left( V +V_A\right) \right) \varepsilon^+ \\
        & - \frac{1}{2} \frac{1}{A} \frac{\partial (AV)}{\partial r}\varepsilon^+ + R_+ + D_+
    \end{aligned}
    \end{equation}
    \begin{equation}\label{eq:epsilon-}
    \begin{aligned}
        \frac{\partial \varepsilon^-}{\partial t} = & - (V-V_A) \frac{\partial \varepsilon^-}{\partial r} - \frac{1}{A} \frac{\partial }{\partial r} \left(A \left( V -V_A\right) \right) \varepsilon^- \\
        & - \frac{1}{2} \frac{1}{A} \frac{\partial (AV)}{\partial r}\varepsilon^- + R_- + D_-
    \end{aligned}
    \end{equation}
\end{subequations}
with $P = P_i+P_e$ and $\varepsilon = \varepsilon^+ + \varepsilon^-$. Here, $B(r)$, $\rho(r)$, $V(r)$, $P_i(r)$, $P_e(r)$ are the radial magnetic field, plasma density, radial solar wind speed, ion thermal pressure, and electron thermal pressure respectively. $\gamma_{i,e}$ are the polytropic indices for ions and electrons respectively. We consider a spherically symmetric radial flux tube such that $A(r)$ is the cross section area of the tube and $A_0$ is the cross section area at the inner boundary. $\varepsilon^\pm = \frac{1}{4} \rho \left| z^\pm \right|^2 $ with $z^\pm$ being the two Els\"asser variables. Thus $\varepsilon^\pm$ represent the energy densities (per volume) of the outward and inward propagating Alfv\'en waves. $V_A = B / \sqrt{\mu_0 \rho}$ is the radial Alfv\'en speed. $D_{\pm}$ are the dissipation rates of the two wave populations due to the nonlinear energy cascade, and $R_{\pm}$ represent the reflection of the waves due to the inhomogeneity of the background plasma. 
$Q_{i}$ and $Q_e$ are the heating terms for ions and electrons, and each of them consist of three components such that $Q_{i} = Q_{h,i} + Q_{w,i} + Q_{c,i}$ and $Q_e = Q_{h,e} + Q_{w,e} + Q_{c,e}$. Here $Q_{h,i}$ and $Q_{h,e}$ are the ad-hoc heating terms that are significant only at very low altitudes. $Q_{w,i}$ and $Q_{w,e}$ are the heating of ions and electrons by the wave dissipation. $Q_{c,i}$ and $Q_{c,e}$ are the heating terms caused by collisionless electron heat conduction \citep{hollweg1976collisionless}.


\begin{figure}
    \centering
    \includegraphics[width=\hsize]{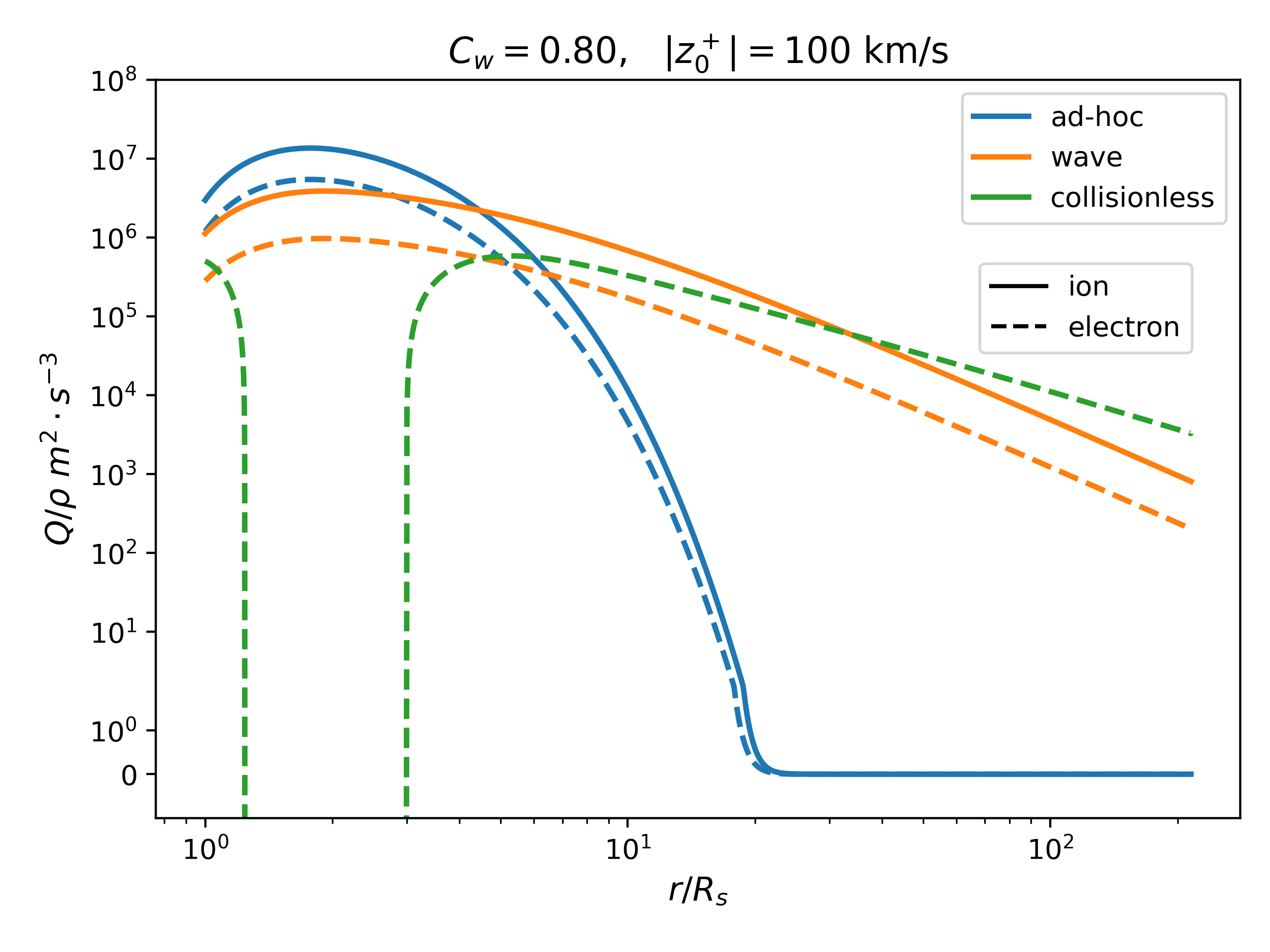}
    \caption{Radial profiles of different heating terms (per unit mass) in the run with $C_w=0.8$ and $| z^+ |=100$ km/s at the inner boundary. Blue curves are the ad-hoc heating, orange curves are the wave heating, and green curve is the collisionless heat conduction. Solid curves are heating of ions and dashed curves are heating of electrons. The minimum of the collisionless heating is around $-6.1\times 10^{5}$ $m^2 \cdot s^{-3}$ at $r=1.85R_s$. }
    \label{fig:heating_terms}
\end{figure}

A comment on the treatment of the Alfv\'en wave equations and their coupling to the solar wind is in order. Though this model has been used before, it represents a drastic simplification of the true problem, as it writes the evolution equations directly in terms of the separate energy densities of outward and inward modes rather than the general second order moment of the fluctuating fields. 
The latter would imply at least four equations rather than the two for the fluctuating energies, and a generalized Reynolds stress in the solar wind momentum equation rather than the simple fluctuating magnetic pressure of equation (\ref{momentum}). 
The still unresolved difficulties of this model have been discussed in, e.g., \cite{velli1993propagation}, while the approximation used above is effective in the limit of small reflection - necessary to trigger nonlinear interactions but not large enough to require the full second order moments - that is satisfied by all except the lowest frequency fluctuations (corresponding to periods of several hours to days).

\begin{figure*}
    \centering
    \includegraphics[width=\hsize]{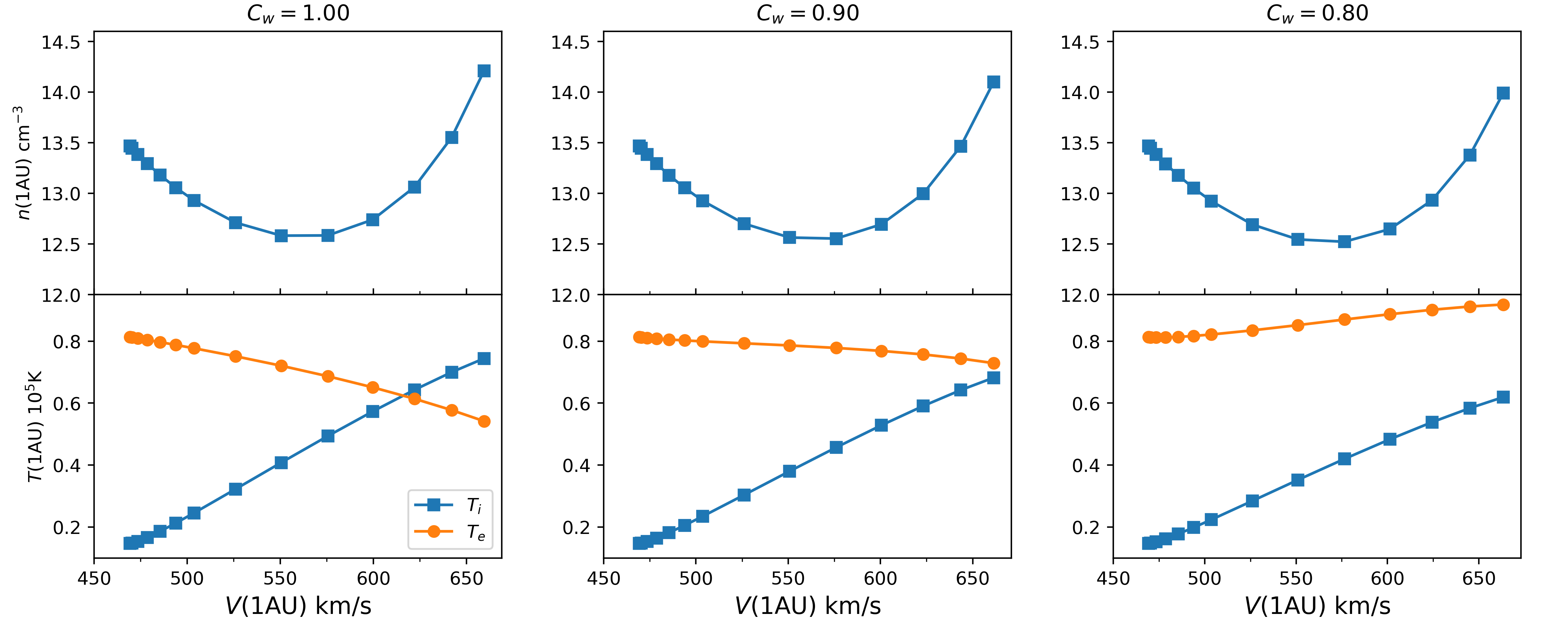}
    \caption{Plasma number density (top), ion temperature (bottom blue), and electron temperature (bottom orange) as functions of the solar wind speed at 1 AU. From left to right panels are runs corresponding to $C_w=$1, 0.9, and 0.8 respectively.}
    \label{fig:quantity_1au_on_V}
\end{figure*}

We use a superradially expanding flux tube \citep{verdini2009turbulence,lionello2014validating}: $A(r) = f(r) r^2$ with
\begin{equation}
    f(r) = \frac{f_{m} + f_1 \exp \left(-(r-r_{exp})/\sigma_{exp}\right) }{ 
             1+ \exp(-(r-r_{exp})/\sigma_{exp})}
\end{equation}
and $f_1 = 1 - (f_m - 1) \exp((r_s-r_{exp})/\sigma_{exp})$. Here $r_s$ is the solar radius, $r_{exp} = 1.31r_s$, $\sigma_{exp} = 0.51 r_s$. The above expression leads  to $f(r_s) = 1$ and $f(+\infty) = f_m$. If $f_m = 1$ we get $f(r) \equiv 1$, which is a radially expanding flux tube. In this study, we set $f_m = 4$, which is a typical value for fast solar wind \citep{wang1990solar}. The polytropic indices are $\gamma_i = \gamma_e = 5/3$ (adiabatic index) so that both the species cool as $T_{i,e} \propto r^{-4/3}$ without other heating terms. 

The first three terms on the right-hand-side of equations (\ref{eq:epsilon+} \& \ref{eq:epsilon-}) correspond to the Wentzel-Kramers-Brillouin (WKB) evolution of the wave amplitudes \citep{alazraki1971solar,belcher1971alfvenic,hollweg1974transverse}. The reflection term is written as
\begin{equation}
    R_\pm = C_R \times \left| (V \mp V_A) \frac{\partial}{\partial r}\ln \sqrt{\rho} \right| \varepsilon^\mp
\end{equation}
with $C_R = 0.1$ being a constant coefficient. The nonlinear dissipation is
\begin{equation}
    D_\pm = - \frac{1}{8}\rho \frac{\left|z^\mp\right| \left|z^\pm \right|^2}{\lambda} =- \frac{\sqrt{\varepsilon^\mp} \varepsilon^\pm}{\sqrt{\rho} \lambda}
\end{equation}
where $\lambda(r)$ is the perpendicular correlation length \citep{reville2020role} and is modeled as
\begin{equation}
    \lambda (r) = \lambda_0 \sqrt{\frac{A(r)}{A_0}}
\end{equation}
In this study, we set $\lambda_0 = 6\times 10^7 m$, similar to the typical size of large supergranules \citep{verdini2007alfven,verdini2009turbulence,reville2020role}. Since the dissipated wave energy heats the protons and electrons, we have
\begin{equation}
    Q_{w,i} + Q_{w,e} = -(D_+ + D_-).
\end{equation}
In the model, a free parameter $C_w \in [0,1]$ controls the portion of the dissipated wave energy that heats the ions:
\begin{equation}
    Q_{w,i} = -C_w(D_+ + D_-),  Q_{w,e} = -(1-C_w)(D_+ + D_-).
\end{equation}
The ad-hoc heating is
\begin{equation}
    Q_{h,(i,e)}= Q_{0,(i,e)} \frac{A_0}{A} \exp \left( - \frac{r-r_s}{r_s}  \right)
\end{equation}
and we set $Q_{0,i} =5  \times 10^{-7} J \cdot m^{-3} \cdot s^{-1}$ and $Q_{0,e} =2 \times 10^{-7} J \cdot m^{-3} \cdot s^{-1}$. These terms represent contributions from processes such as the nanoflares \citep[e.g.][]{cargill2004nanoflare} that are important in the low corona. We choose $Q_{0,e} < Q_{0,i}$ because remote-sensing observations reveal that the electron temperature is smaller than the proton temperature in the coronal holes \citep{cranmer2009coronal}. 
The collisionless heat conduction term writes as 
\begin{equation}
    Q_{c,i} = 0, \, Q_{c,e} = -  \frac{1}{A}  \frac{\partial}{\partial r} \left(A  q_c \right)
\end{equation}
where $q_c = \frac{3}{2} P_e V$ \citep{hollweg1976collisionless}. Note that the collisionless heat conduction takes effect for electrons only.

\begin{figure*}
    \centering
    \includegraphics[width=\hsize]{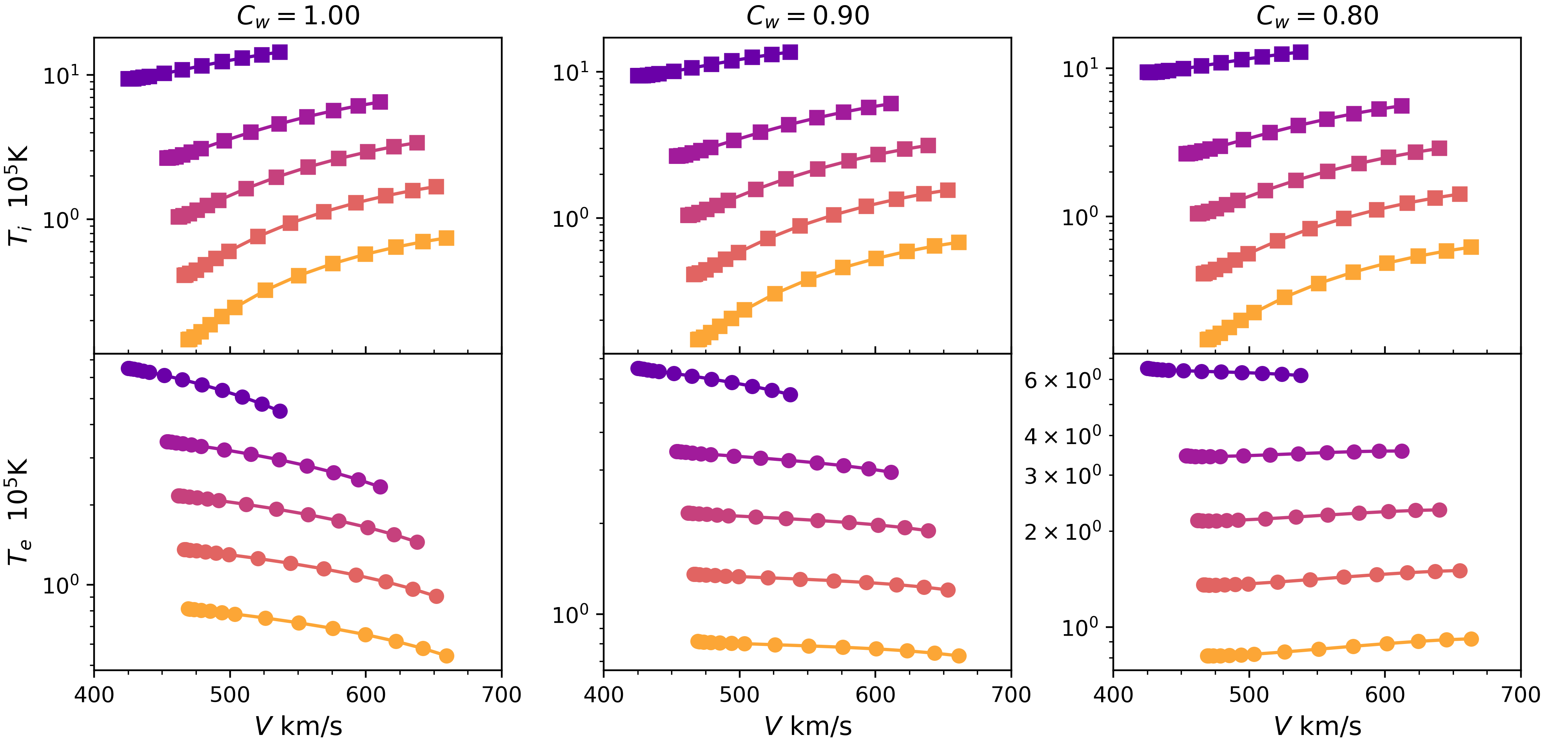}
    \caption{$T_i-V$ (top) and $T_e-V$ (bottom) correlations at different radial distances to the Sun in runs with $C_w=$1 (left), 0.9 (middle), and 0.8 (right). In each panel, from dark to light colors correspond to $r=$10, 25, 50, 100, and  215 solar radii.}
    \label{fig:quantity_on_V_diff_r}
\end{figure*}

The simulation domain is $r\in[1,215]r_s$ with nonuniform grid. The spatial resolution is $\Delta x=0.001r_s$ at the inner boundary and $\Delta x = 0.1 r_s$ at the outer boundary, and the total number of grid points is $N=2934$. Dirichlet boundary conditions are imposed for $(B,\rho,P_i,P_e,\varepsilon^+)$ at the inner boundary and the outer boundary is open so the wind and waves can propagate out of the domain freely. We note that no inner boundary conditions are needed for $V$ and $\varepsilon^-$ because the sonic point and Alfv\'en point implicitly impose two constraints for them \citep{parker1958dynamics,barkhudarov1991alfven,velli1993propagation}. In all the simulations, we set $B = 5G$, $n=1\times10^8 cm^{-3}$, $T_i=T_e=1 MK$ at the inner boundary where $n$ is the number density of the plasma, $T_i$ and $T_e$ are the ion temperature and electron temperature respectively. We carry out five sets of simulations with $C_w=[1, 0.95, 0.9, 0.85, 0.8]$. For each $C_w$, we do a series of runs with varying $\varepsilon^+ (r_s)$ such that $\left|z^+ \right| = [0,5,10,15,20,25,30,40,50,60,70,80,90,100]$ km/s at the inner boundary. We run each simulation until all the fields reach a stationary state ($\partial_t = 0$) and acquire the radial profiles of the fields.

\subsection{Results}
In Figure \ref{fig:radial_profile}, we plot the radial profiles of various quantities in two sets of runs. The left column shows runs with $C_w=1$ and the right column shows runs with $C_w=0.8$. From top to bottom rows are solar wind speed, 
plasma number density, ion temperature, and electron temperature. In each panel, dark to light colors correspond to runs with increasing values of the inner boundary wave amplitude from 0 to 100 km/s. Larger wave energy input leads to a higher solar wind speed because of a stronger wave pressure gradient and more heating of the plasma. The wave amplitude does not change the density profile much. The left column clearly shows that larger wave amplitude leads to higher ion temperature and lower electron temperature. For $C_w=0.8$, the ion temperature still increases with the wave amplitude, while the behavior of electron temperature is more complicated. Close to the Sun ($r \lesssim 20 r_s$),  electron temperature decreases with wave amplitude. Further away from the Sun, electron temperature increases with the wave amplitude. 
Figure \ref{fig:heating_terms} shows contributions to ion heating (solid curves) and electron heating (dashed curves) per unit mass, i.e. $Q/\rho$, by different mechanisms in the run with $C_w=0.8$ and $|z^+|=100$ km/s at the inner boundary. Blue curves are the ad-hoc heating, orange curves are the wave heating, and the green curve is the collisionless electron thermal conduction. The wave heating is weaker than the ad-hoc heating close to the Sun, but the radial extent of significant wave heating is larger than the ad-hoc heating. In addition, the collisionless thermal conduction is comparable or even larger than wave heating far away from the Sun.

\begin{figure}
    \centering
    \includegraphics[width=\hsize]{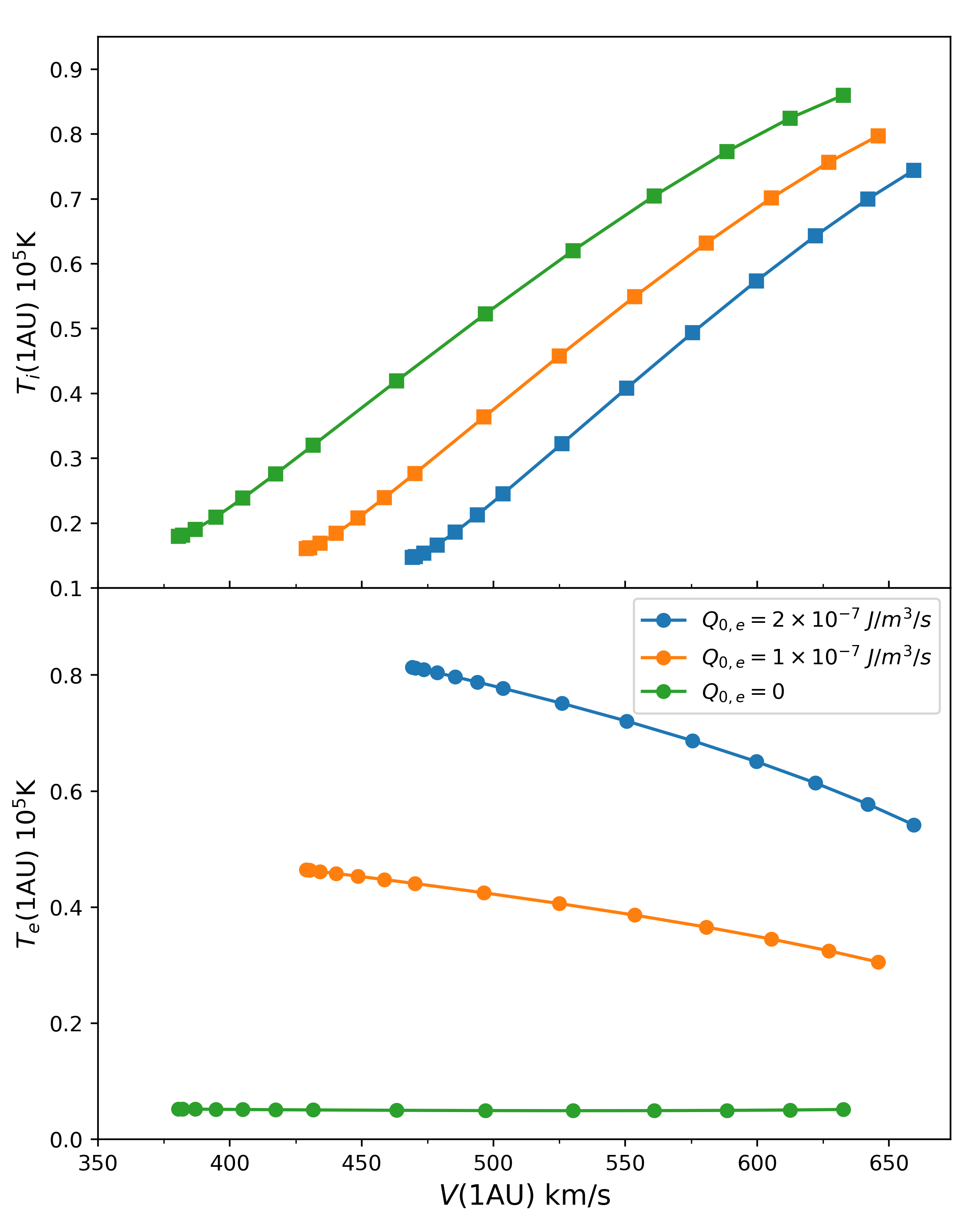}
    \caption{$T_i-V$ (top) and $T_e-V$ (bottom) correlations at 1AU for three sets of runs with different amplitudes of electron ad-hoc heating. Blue curves are $Q_{0,e}=2 \times 10^{-7}$ $J/m^3/s$, orange curves are $Q_{0,e}= 1 \times 10^{-7}$ $J/m^3/s$, and green curves are $Q_{0,e} = 0$. All the runs have $C_w=1$. }
    \label{fig:Ti_Te_diff_C_AH_electron}
\end{figure}

In Figure \ref{fig:quantity_1au_on_V}, we show how the plasma number density (top panel), ion temperature (bottom blue) and electron temperature (bottom orange) vary with the solar wind speed at 1 AU in runs with different $C_w$. From left to right columns are $C_w=1$, $0.9$, and $0.8$ respectively. The behavior of plasma density does not depend on $C_w$ as how the dissipated wave energy is distributed among ions and electrons does not affect the radial profile of solar wind speed or the density much. As we increase the wave amplitude, the density drops at first and then starts to increase, though only with small variation ($n$ varies between 12.5 and 14.5 cm$^{-3}$). Since the mass conservation law gives $n(r) = n_0 \times (A_0 V_0 / A(r) V(r))$, as the wave amplitude increases, if $V_0$ increases slower than $V(r)$, $n(r)$ has an anti-correlation with $V(r)$, and vice versa. Hence, the small variation of density with wave amplitude indicates that the waves modify $V_0$ and $V(r)$ with similar proportions.
Similar to the density, ion temperature is not modified by $C_w$ significantly, either. On the contrary, the electron temperature is quite sensitive to $C_w$. As $C_w$ decreases, the negative $T_e-V$ correlation gradually turns to positive, consistent with what is shown by Figure \ref{fig:radial_profile}.

In Figure \ref{fig:quantity_on_V_diff_r}, we show $T_i-V$ (top row) and $T_e-V$ (bottom row) at different radial distances to the Sun for runs with $C_w=1$ (left), $C_w=0.9$ (middle), and $C_w=0.8$ (right) respectively. In each panel, from dark to light colors correspond to $r=10$, $25$, $50$, $100$, and $215$ solar radii. For $C_w=1$, positive $T_i-V$ correlation and negative $T_e-V$ correlation are well established at very close distance to the Sun and maintained as the wind propagates. For $C_w=0.9$ and $C_w=0.8$, the ion temperature is not modified much, while the $T_e-V$ correlation evolves as the solar wind propagates. Close to the Sun, negative $T_e-V$ correlation is produced, while as $r$ increases, $T_e-V$ correlation gradually turns positive. This trend is similar to in-situ measurements by multiple satellites \citep{maksimovic2020anticorrelation}.

\section{Discussion}\label{sec:discuss}
The positive $T_i-V$ correlation is easy to understand: With more wave energy injected from the inner boundary, the solar wind speed increases because of larger wave pressure and larger thermal pressure gradient. Meanwhile, because most of the dissipated wave energy heats the ions, the ion temperature also increases, resulting in a positive $T_i-V$ correlation. The cause of negative $T_e-V$ correlation is more complicated. If we consider the most simple case where the electron fluid is polytropic such that $T_e(r) = T_{e0} \times \left( \rho_0 / \rho(r) \right)^{\gamma - 1} $, the $T_e-V$ relation should be similar to $n-V$ relation. However, the left column of Figure \ref{fig:quantity_1au_on_V} shows that even in the large-$V$ regime where the density increases with $V$, $T_e$ still decreases with $V$. This indicates that the ad-hoc heating may play an important role in forming the $T_e-V$ anti-correlation. By observing the equation for electron pressure (equation (\ref{eq:pe})), we see that, in a stationary state ($\partial_t = 0$) and without Alfv\'en wave heating of the electrons, the radial gradient of the electron pressure can be written as
\begin{equation}
    \frac{\partial P_e}{\partial r} = C_1 P_e \frac{1}{A} \frac{\partial A}{\partial r} + C_2 P_e \frac{1}{V} \frac{\partial V}{\partial r} + C_3 \frac{Q_{h,e}}{V}
\end{equation}
where $C_1$, $C_2$, and $C_3$ are constants depending on $\gamma_e$ and the collisionless heat conduction strength. 
Close to the inner boundary, the dominating term is the ad-hoc heating term (Figure \ref{fig:heating_terms}). Because $Q_{h,e}(r)$ is a given function of $r$, the contribution of this term to the increment of pressure is inversely proportional to the solar wind speed. That is to say, the faster the plasma is ejected, the less internal energy it gains during its propagation. 
This is why a stronger Alfv\'en wave injection leads to a lower electron temperature at close distances to the Sun.
In Figure \ref{fig:Ti_Te_diff_C_AH_electron}, we show $T_i-V$ (top) and $T_e-V$ (bottom) relations at 1AU for three sets of runs with $C_w=1$ and varying amplitudes of electron ad-hoc heating. The blue curves correspond to $Q_{0,e}=2 \times 10^{-7}$ $J/m^3/s$, the orange curves correspond to $Q_{0,e}=1 \times 10^{-7}$ $J/m^3/s$, and the green curves correspond to $Q_{0,e}=0$. One can see that, as we decrease  $Q_{0,e}$, the negative $T_e-V$ correlation gradually vanishes, implying that the ad-hoc heating is necessary for the negative $T_e-V$ correlation.
However, if the electrons gain a portion of the wave energy during the solar wind expansion, the anti-correlation between $T_e$ and $V$ is gradually destroyed, because further away from the Sun the contribution of the ad-hoc heating gradually becomes less important compared with the contribution of the wave dissipation (Figure \ref{fig:heating_terms}). 
This explains why there is a radial evolution of the $T_e-V$ relation. 
Based on this scenario, the in-situ observations (Figure \ref{fig:insitu_observation}) indicate that in the slow solar wind, electrons get more heating during the solar wind propagation compared with electrons in the fast solar wind, leading to different $T_e-V$ correlations for slow and fast streams (right column of Figure \ref{fig:insitu_observation}). The underlying mechanisms, however, need further studies.


\section{Conclusion}\label{sec:conclusion}
Through a 1D Alfv\'en-wave-driven solar wind model with different ion and electron temperatures, we have successfully reproduced two important features of the solar wind, namely a positive correlation between the ion temperature ($T_i$) and solar wind speed ($V$) and a negative correlation between the electron temperature ($T_e$) and solar wind speed. 
In our simulations, the different $T_i-V$ and $T_e-V$ relations are a result of the fact that most of the dissipated Alfv\'en wave energy heats ions instead of electrons \citep[e.g.][]{arzamasskiy2019hybrid,squire2022high,bacchini2022kinetic}, making electron heating  close to the Sun by the ad-hoc heating term, which represents mechanisms such as magnetic reconnection, less efficient due to faster wind speed. 
With a small but finite portion of the dissipated wave energy heating the electrons, the simulations also reproduce the observed radial evolution of the $T_e-V$ relation, i.e., the initially negative correlation gradually turns into a positive one \citep{maksimovic2020anticorrelation}, because the contribution of the ad-hoc heating term gradually becomes negligible compared with the Alfv\'en wave heating as the wind propagates. 

We note that the model used here is not fully self-consistent and some important characteristics of the observed solar wind are missing in the model results. 
First, the temperature evolution given by the model only qualitatively, but not completely quantitatively, agrees with the observations. 
The radial decay rate ($\alpha$ in $T_e \propto r^{-\alpha}$) of the electron temperature is large and does not vary much with solar wind speed. In the bottom-right panel of Figure \ref{fig:radial_profile}, $\alpha$ changes from $\sim 0.67$ to $\sim 0.65$ as the wind speed at 1AU increases from $470$km/s to $660$km/s. 
In contrast, $\alpha$ varies between $\sim 0.4$ and $\sim 0.2$ as wind speed changes from $400$km/s to $600$km/s as estimated using HELIOS data \citep{maksimovic2020anticorrelation}. 
This difference implies that other mechanisms, e.g. the ambipolar electric field \citep{boldyrev2020electron}, omitted in our model play an important role in local electron heating. 
Second, the density only changes slightly among runs with different wave amplitudes (Figure \ref{fig:radial_profile}) and thus density does not vary with the wind speed significantly (top row of Figure \ref{fig:quantity_1au_on_V}). However, it is well known that faster solar wind is generally less dense than slower solar wind and the mass flux only moderately depends on the solar wind speed \citep{wang2010relative}. Both the inner boundary plasma density and the expansion factor can heavily affect the solar wind density, but they are both constant in the current study. 
A thorough parametric study in which all these parameters are treated as variables is necessary and will be conducted in the future. 
Hence, the current study serves as a demonstration of how the Alfv\'en waves can contribute to the observed $T-V$ correlations in the solar wind, while other mechanisms \citep[e.g.][]{fisk2003acceleration} may still be important and should be incorporated in a complete description of the system.

\begin{acknowledgments}
This work is supported by NASA HTMS 80NSSC20K1275 and the NASA Parker Solar Probe Observatory Scientist grant NNX15AF34G. The instruments of PSP were designed and developed under NASA contract NNN06AA01C. We thank Dr. Kun Zhang for many useful suggestions.
\end{acknowledgments}

%

\vspace{5mm}


\software{Matplotlib \citep{Hunter2007Matplotlib}}






\bibliography{references}{}
\bibliographystyle{aasjournal}



\end{CJK*}
\end{document}